\documentclass[aps,prd,twocolumn,superscriptaddress,10pt]{revtex4-2}

\usepackage{amsfonts}
\usepackage{amsmath}
\usepackage{amssymb}

\usepackage{graphicx}
\usepackage[dvipsnames]{xcolor}
\definecolor{rossocorsa}{rgb}{0.83, 0.0, 0.0}
\definecolor{bleudefrance}{rgb}{0.19, 0.55, 0.91}

\usepackage{hyperref}
\hypersetup{
    colorlinks,
    citecolor=bleudefrance,
    filecolor=bleudefrance,
    linkcolor=rossocorsa,
    urlcolor=bleudefrance
}
\usepackage{cleveref}
\usepackage{footmisc}

\newcommand{\iu}{\text{i}}
\newcommand{\Ginv}{G^{(\text{inv})}}
\newcommand{\e}[1]{\text{e}^{#1}}
\newcommand{\G}{\mathcal{G}}
\newcommand{\Op}{\mathcal{O}}
\newcommand{\diff}{\text{d}}
\newcommand{\dDisc}[1]{\text{dDisc}\left[#1\right]}
\newcommand{\dDisct}[1]{\text{dDisc}_{t}\left[#1\right]}
\newcommand{\dDiscu}[1]{\text{dDisc}_{u}\left[#1\right]}

\newcommand{\wb}{\bar w}
\newcommand{\zb}{\bar z}
\newcommand{\rz}{\rho_z}
\newcommand{\rw}{\rho_w}
\newcommand{\rzb}{\bar\rho_z}
\newcommand{\rwb}{\bar\rho_w}

\newcommand{\rt}{\mathrm{t}}
\newcommand{\ru}{\mathrm{u}}

\begin{document}


\title{Conformal dispersion relation for mixed correlators}
\author{Dean Carmi}
\email{deancarmi1@gmail.com}
\affiliation{Department of Physics and Haifa Research Center for Theoretical Physics and Astrophysics, University of Haifa, Haifa 31905, Israel}

\author{Javier Moreno} 
\email{fjaviermoreno@udec.cl}
\affiliation{Departamento de Física, Universidad de Concepción, Casilla, 160-C, Concepción, Chile}
\affiliation{Departament de Física Quàntica i Astrofísica, Institut de Ciències del Cosmos, Universitat de Barcelona, Martí i Franquès 1, E-08028 Barcelona, Spain}

\author{Shimon Sukholuski}
\email{ssm@technion.ac.il}
\affiliation{Department of Physics and Haifa Research Center for Theoretical Physics and Astrophysics, University of Haifa, Haifa 31905, Israel}

\begin{abstract}
Dispersion relations are nonperturbative formulas that relate the ultraviolet and infrared behavior of an observable  with wide-ranging applications applications in linear response theory, quantum field theory scattering amplitudes, and conformal correlators.
We derive a position-space dispersion relation for scalar four-point mixed correlation functions in an arbitrary conformal field theory. This formula expresses the correlator in terms of its integrated double discontinuity times a kinematic kernel. The kernel is analytically computed, and expressed in a remarkably simple form as a two-variable Appell function. The dispersion kernel is found by solving a coupled partial differential equation that the kernel obeys. Numerical checks of the dispersion relation are successfully performed for generalized free field correlators. Finally, we show that our position-space dispersion relation is equivalent to a Cauchy-type dispersion relation of the Mellin amplitude of the correlator.
\end{abstract}

\maketitle
\noindent \textit{Introduction}. Dispersion relations (DRs) are mathematical tools that reconstruct complex functions from their discontinuities. Their first major application in physics was by Kramers and Kronig, who related the real and imaginary parts of the refractive index as a function of frequency~\cite{de1926theory,Kramers:1928}. There, the imaginary part---which reflects how the material absorbs energy---is enough to determine the full response. This reconstruction relies on causality, which ensures analyticity in the upper half of the complex frequency plane—a key requirement for the validity of DRs. In particular, a Cauchy-type DR expresses the function $\mathcal M(s)$ in terms of its discontinuity across the branch cut.
\begin{equation}\label{eq:cauchy}
\mathcal{M}(s) = \frac{1}{2\pi \iu}\int_{\text{cut}} \diff s’ \frac{\text{Disc}[\mathcal{M}(s’)]}{s’-s}\,.
\end{equation}

An important application of DRs arose in the 1950s with the S-matrix bootstrap program, which aimed to determine the spectrum and interactions of strongly coupled hadrons from first principles like analyticity and high-energy behavior---see e.g.~\cite{Mandelstam:1958xc,Chew:1960iv,Martin:1969ina,Eden:1966dnq,screaton1961dispersion}. Recently, the S-matrix bootstrap has been revived through modern analytic and numerical methods~\cite{Paulos:2016fap,Paulos:2017fhb,Kruczenski:2022lot}, as well as insights from the anti-de Sitter/conformal field theory (AdS/CFT) correspondence~\cite{Maldacena:1997re,Gubser:1998bc,Witten:1998qj}.

In the context of CFT---which describes physical systems at criticality and finds applications ranging from condensed matter physics to string theory—the bootstrap approach has achieved remarkable success~\cite{Rattazzi:2008pe,El-Showk:2012cjh,El-Showk:2014dwa,Simmons-Duffin:2016gjk,Poland:2018epd}. The primary observables in a CFT are correlation functions of local operators, $\langle \mathcal{O}_1 \cdots \mathcal{O}_n \rangle$, where each operator $\mathcal{O}_i(x_i)$ carries a scaling dimension $\Delta_i$. A central goal of the conformal bootstrap is to determine these correlators---or, equivalently---to extract the spectrum of anomalous dimensions (critical exponents) and operator product expansion (OPE) coefficients. A key advantage of this approach is its nonperturbative nature, which allows access to precise results even at strong coupling.

In the numerical conformal bootstrap, bounds on conformal data from four-point functions of identical operators become much tighter when including mixed correlators of nonidentical operators. This often produces small allowed ``islands’’~\cite{Kos:2016ysd,Kos:2015mba,Kos:2014bka}, enabling precise determination of critical exponents, as in the 3$d$ Ising model.

Dispersive techniques entered the conformal bootstrap in~\cite{Caron-Huot:2017vep}, which introduced the double discontinuity and the Lorentzian inversion formula (LIF)---see also~\cite{Simmons-Duffin:2017nub,Kravchuk:2018htv}. A position-space DR for four-point functions with equal dimensions was later derived in~\cite{Carmi:2019cub}, while Mellin-space versions appeared in~\cite{Penedones:2019tng}. Their equivalence was established in~\cite{Caron-Huot:2020adz}.  These developments highlight the power of DRs in extracting nonperturbative conformal data from analyticity and symmetry alone, making them a compelling and elegant tool in the ongoing effort to solve CFTs with a nonperturbative approach.

In this work, we derive a position-space conformal DR for scalar four-point mixed correlators, generalizing~\cite{Carmi:2019cub} to unequal scaling dimensions. We also show that our formula is equivalent to a Cauchy-type DR for conformal correlators in Mellin space. Finally, we perform successful checks of the formula, and discuss potential applications of the conformal DR for bounding conformal data via the mixed-correlator dispersive conformal bootstrap.

\noindent\textit{Lorentz inversion formula}. Consider a CFT in $d\geq 2$ space time dimensions, and the four-point correlation functions of operators $\Op_i$. The latter can be written in terms of two cross ratios $z$, $\bar z$ defined through the relations $U=z\bar z=x_{12}^2x_{34}^2/(x_{13}^2x_{24}^2)$, $V=(1-z)(1-\bar z)=x_{23}^2x_{14}^2/(x_{13}^2x_{24}^2)$, with $x_{ij}=x_i-x_j$. Then the four-point function is
\begin{equation}\label{eq:4point}
\langle\Op_1\Op_2\Op_3\Op_4\rangle=\frac{\left(x_{14}^{2}/x_{24}^{2}\right)^{a}\left(x_{14}^{2}/x_{13}^{2}\right)^{b}}{\left(x_{12}^{2}\right)^{\frac{\Delta_1+\Delta_2}{2}}\left(x_{34}^{2}\right)^{\frac{\Delta_3+\Delta_4}{2}}}\G(z,\zb)\, ,
\end{equation}
where $a\equiv(\Delta_2-\Delta_1)/2$ and $b\equiv (\Delta_3-\Delta_4)/2$, and $\G(z,\zb)$ is 
 a function that depends on the particular dynamics of the CFT. We will frequently use radial coordinates~\cite{Hogervorst:2013sma}, defined as $\rz\equiv (1-\sqrt{1-z})/(1+\sqrt{1-z})$, $\rzb\equiv (1-\sqrt{1-\bar z})/(1+\sqrt{1-\bar z})$.

The OPE implies a principle series expansion of $\G(z,\zb)$ in terms of conformal partial waves $F_{J,\Delta}$:
\begin{equation} \label{eq:ope4}
\G(z,\zb)=1_{12}1_{34}+\sum_{J=0}^\infty\int_{\frac{d}{2}-\iu\infty}^{\frac{d}{2}+\iu\infty}\frac{\diff\Delta}{2\pi \iu}c_{J,\Delta}F_{J,\Delta}(z,\bar z)\, , 
\end{equation}
where $F_{J,\Delta}$ are a shadow symmetric combination of the conformal blocks $G_{J,\Delta}$. The coefficients of the expansion, $c_{J,\Delta}$, are called OPE functions. The location of poles of $c_{J,\Delta}$ in the complex plane gives the spectrum of scaling dimensions of the theory.

The LIF introduced in~\cite{Caron-Huot:2017vep} inverts the conformal block expansion~\eqref{eq:ope4}, by extracting the OPE function $c_{J,\Delta}$ from the double discontinuity of $\G(z,\zb)$
\begin{align}\label{eq:invc}
c_{J,\Delta}^t&=\frac{\kappa}{4}\int_0^1\diff w\diff \wb\,\mu\, \Ginv_{J,\Delta}(w,\bar w)\dDisct{\G(w,\wb)}\,,
\end{align}
where $c_{J,\Delta}=c_{J,\Delta}^t+(-1)^Jc^u_{J,\Delta}$. The ``inverted block" is defined from the conformal block as $\Ginv_{J,\Delta}(w,\bar w)\equiv G_{\Delta+1-d,J+d-1}(w,\wb)$, while $\mu\equiv\mu_{(d)}^{(a,b)}(w,\wb)=\left|(w-\wb)/(w\wb)\right|^{d-2}\left[(1-w)(1-\wb)\right]^{a+b}/(w\wb)^2$ and $\kappa\equiv \kappa_{J+\Delta}=\Gamma_{\beta/2-a}\Gamma_{\beta/2+a}\Gamma_{\beta/2-b}\Gamma_{\beta/2+b}/(2\pi^2\Gamma_{\beta-1}\Gamma_\beta)$ \footnote{We are employing the notation $\Gamma_x\equiv\Gamma(x)$ for the gamma function.}. The double discontinuity around $\zb=1$ is:%
\begin{align}
&\dDisct{\G(z,\zb)}=\cos[\pi(a+b)]\G(z,\zb)\notag\\
&-\frac{1}{2}\e{\iu\pi (a+b)}\G\left(z,\zb^\circlearrowleft\right)\,
-\frac{1}{2}\e{-\iu\pi (a+b)}\G\left(z,\zb^\circlearrowright\right)\,.\label{eq:disct}
\end{align}

 \noindent \textit{Conformal dispersion relation}. We aim to derive a  \textit{conformal DR} that expresses the four-point correlator $\G(z,\bar z)$ as an integral over its (double) discontinuities, in the spirit of~\eqref{eq:cauchy}. A method to derive such a DR was devised in~\cite{Carmi:2019cub}. It consists of combining the LIF~\eqref{eq:invc}, inside the OPE expansion~\eqref{eq:ope4}. Replacing the order of integrations gives the conformal DR,
\begin{equation} \label{eq:Gt}  \boxed{ \mathcal{G}^t(z,\bar{z})=\int\diff w\diff \wb\,  K(z, \bar z,w,\bar w)\,\dDisc{\G(w,\wb)}}
\end{equation}
and $G(z,\zb)=\G^t(z,\zb)+\G^u(z,\zb)$.
The kernel $K\equiv K^{(a,b)}(z, \bar z,w,\bar w)$ reads 
\begin{equation}\label{eq:Kpre}
K=\frac{\mu}{8\pi\iu}\sum_{J=0}^\infty\int \diff\Delta  \kappa_{J+\Delta} F_{J,\Delta}(z,\bar z)\Ginv_{J,\Delta}(w,\bar w)\, .
\end{equation}
The conformal DR~\eqref{eq:Gt} is a remarkable nonperturbative formula that bootstraps the correlator $\G(z,\zb)$.
However in order to apply it efficiently, the kernel $K$ in~\eqref{eq:Kpre} needs to be computed analytically. We can show numerically that $K$ is independent of the space-time dimension $d$, just as the Cauchy-type DR~\eqref{eq:cauchy} is. This intriguing fact arises because the DR should hold for any function $\G(z,\bar z)$ with analytic properties as the four-point correlator, and these analytic properties do not depend on $d$. Furthermore, the kernel can be written as a sum of two terms, i.e.,
\begin{align}\label{eq:Kdecomp}
    &K(z,\bar z,w,\bar w)=
    K_B(z,\bar z,w,\bar w)\theta(\rz\rzb\rwb-\rw)\notag\\
    &+K_C(z,\bar z,w,\bar w)\frac{\diff \rw}{\diff w}\delta(\rw-\rz\rzb\rwb)\, ,
\end{align}
where $\delta$ is the Dirac delta function, and the unit step function $\theta$ implies that the integration region in the DR~\eqref{eq:Gt} has support only for 
$\rw < \rz\rzb\rwb$.

The contact term $K_C$ was computed~\cite{Carmi:2019cub} for general external scaling dimensions $\Delta_i$, i.e., for any $a$ and $b$ while the bulk term $K_B$ was computed explicitly only for $a=b=0$, see \footnote{The bulk kernel and contact kernel are given by~\cite{Carmi:2019cub}:
\begin{align}
K_B^{(a=0,b=0)}&=\left(\frac{z \bar z}{w \bar w}\right)^{3/2}\frac{(w-\bar w)(w^{-1}+\bar w^{-1}+z^{-1}+\bar z^{-1}-2)}
{\left(\left(1-z\right)\left(1-\bar z\right)\left(1-w\right)\left(1-\bar w\right)\right)^{3/4}}
\frac{x^{\frac{3}{2}}}{64\pi}{}_2F_1\left(\frac{1}{2},\frac{3}{2},2,1-x\right)\,\label{eq:KBa0b0} ,\\
K_C^{(a,b)}&= \left( \frac{(1-w)(1-\bar w)}{(1-z)(1-\bar z)}\right)^{\frac{a+b}{2}}\frac{4}{\pi}\frac{1}{\bar w^2}\left(\frac{1-\rz^2\rzb^2\rwb^2}{(1-\rz^2)(1-\rzb^2)(1-\rwb^2)}\right)^{1/2}\frac{1-\rz\rzb\rwb^2}{(1-\rz\rwb)(1-\rzb\rwb)}\,,\label{eq:KCa0b0}
\end{align}
where ${}_2F_1$ is the hypergeometric function, and $x$ and $y$ are defined in~\eqref{eq:x} and~\eqref{eq:y}}. In the present work we close this gap, and compute $K_B$ for any $a$ and $b$, i.e., for general mixed correlation functions.

\noindent \textit{Computing the kernel}: We present two complementary methods to compute the bulk kernel $K_B$ in~\eqref{eq:Kdecomp}.

\noindent \underline{I. Series representation~\cite{Carmi:2019cub}:}
\noindent Since the kernel is independent of the dimension, we set $d=2$ in~\eqref{eq:Kpre}, in which we have explicit expressions for the conformal blocks: $G_{J,\Delta}(z,\bar z)=\left[k_{\Delta-J}(z)k_{\Delta+J}(\bar z)+(z \leftrightarrow \bar z)\right]/(1+\delta_{J,0})$ where $k_\beta(z)\equiv z^{\beta/2} {}_2F_1 (\beta/2+a,\beta/2+b,\beta,z)$. Now we perform the $\Delta$ integral in~\eqref{eq:Kpre} using the residue theorem. There are four towers of poles coming from the $\kappa$ factor, and the $J$ sum can be performed. The contribution of the four towers of poles (see~\cite{Carmi:2019cub}):
\begin{align} \label{eq:KKi}
&K_B(z,\bar z,w,\bar w)=  K^{\left(a,b\right)}_{a}+K^{\left(a,b\right)}_{-a}+K^{\left(b,a\right)}_{b}+K^{\left(b,a\right)}_{-b}\\ \nonumber
&=\mu\,\mathcal{D}_2\left(S_{a}^{(a,b)}+S_{-a}^{(a,b)}+S_{b}^{(b,a)}+S_{-b}^{(b,a)}\right)=\mu\,\mathcal{D}_2 S\,,
\end{align}
where the differential operator $\mathcal{D}_2$ is \footnote{\label{foot1}We define the first-order differential operators $\mathcal D_2\equiv z w/(z-w)\left[(1-w)\partial_w-(1-z)\partial_z\right]+(z\leftrightarrow \zb)-1$, $\mathcal D_4\equiv z\zb(w-\wb)/[(z-\zb)w\wb]\left(zw\left[(1-w)\partial_w-(1-z)\partial_z\right)/(z-w)+\left(z\leftrightarrow\zb\right)\right]$, and the second-order one $\mathcal{D}_{\rho_z}\equiv \rho_z ^2\partial_{\rho_z}^2+2\rho_z ^2\left(2 a+2 b+\rho_z\right)\partial_{\rho_z}/(\rho_z ^2-1)-4 a b \rho_z/(\rho_z +1)^2$.}, and we have:
\begin{widetext}
\begin{align}
S_{a'}^{(a,b)}=\sum_{m=0}^\infty\frac{\sin(2\pi a')}{2\pi m!}\frac{\Gamma_{1+2a'+2m}^2k_{-2m-2a'}(z)k_{-2m-2a'}(\bar z)k_{2m+2a'+2}(w)k_{-2m-2a'}(\bar w)}{\Gamma_{1+2a'+m}\Gamma_{1+a'-b+m}\Gamma_{1+a'+b+m}\sin[\pi(a-b)]\sin[\pi(a+b)]}\,.\label{eq:Si}
\end{align}
\end{widetext}
The sum over $m$ in~\eqref{eq:Si} arises from the sum over the residues of poles in the right-hand side of~\eqref{eq:Kpre}. Expression~\eqref{eq:Si} gives a representation of the bulk kernel in terms of an infinite sum of a product of four hypergeometric functions. Our goal is to obtain an analytic expression for this sum. In~\cite{Carmi:2019cub}, the case $a = b = 0$ was analyzed, where each of the hypergeometric functions $k_\beta$ reduces to a Legendre function. This simplification allowed for the explicit evaluation of the 
sum over the double poles in~\eqref{eq:Kpre}. The bulk kernel $K_B$ is subsequently obtained by applying the differential operator given in~\eqref{eq:KKi}. The resulting expression for the bulk kernel in the case $a = b = 0$ is~\footnotemark[2].

For general mixed correlators with $a, b \neq 0$, an analytic evaluation of the sum in~\eqref{eq:Si} is unknown. We therefore turn to an alternative method for computing the kernel.

\noindent \underline{II. Differential equations:} Repeating the computation for $d = 4$ yields a result analogous to Eqs. \eqref{eq:KKi}--\eqref{eq:Si}. Imposing that the kernel in $d = 4$ matches its counterpart in $d = 2$ then leads to the following two constraints:
\begin{equation}\label{eq:D2D4e}
\mathcal{D}^{(24)}_{(z,\bar z,w ,\bar w)}  S_{a'}^{(a,b)}=0\,,\quad \mathcal{D}^{(24)}_{(\bar w,\bar z,w ,z)}   S_{a'}^{(a,b)}=0\,,
\end{equation}
with $\mathcal{D}^{(24)}_{(z,\bar z,w ,\bar w)} \equiv \mathcal D_2-\mathcal D_4$ and $\mathcal D_4$ defined in \footnotemark[3]. These constraints arise because $S$ and $\mathcal{D}^{(24)}_{(z,\bar z,w ,\bar w)}$ have permutation symmetry in the variables $(z,\bar z,\bar w)$ and $(w,z,\bar z)$, respectively. Applying~\eqref{eq:D2D4e} to the four-variable function $S(z,\bar z, w, \bar w)$, reduces its dependence two variables, namely \footnote{This is because we have
$\mathcal{D}^{(24)}_{(z,\bar z,w ,\bar w)}\big[\sqrt{z\bar z w\bar w}\big] =\mathcal{D}^{(24)}_{(\bar w,\bar z,w ,z)}\big[\sqrt{z\bar z w\bar w}\big] =0$ as well as $\mathcal{D}^{(24)}_{(z,\bar z,w ,\bar w)}\big[f(x,y)\big] =\mathcal{D}^{(24)}_{(\bar w,\bar z,w ,z)}\big[f(x,y)\big] =-f(x,y)$, where $f(x,y)$ is any function of variables $x$ and $y$ in~\eqref{eq:x}-\eqref{eq:y}.}
\begin{equation}\label{eqstilde4}
S_{a'}^{(a,b)}=\frac{\sqrt{z\zb w\wb}}{y^{\frac{1}{2}+a+b}}\tilde S_{a'}^{(a,b)}(x,y)\,,
\end{equation}
where the relation between the two variables, $x$ and $y$ and the radial coordinates is
\begin{align}
x&\equiv\frac{\frac{\rz\rzb\rw\rwb}{(1-\rz\rzb\rw\rwb)}(1-\rz^2)(1-\rzb^2)(1-\rw^2)(1-\rwb^2)}{(\rzb\rwb - \rz\rw)(\rz\rwb-\rzb\rw)(\rz\rzb-\rw\rwb)}\, ,\label{eq:x}\\
y&\equiv\frac{(1-\rz)(1-\rzb)(1-\rw)(1-\rwb)}{(1+\rz)(1+\rzb)(1+\rw)(1+\rwb)}\,,\label{eq:y} 
\end{align}
Note that the dispersion kernel $K(z,\bar z, w, \bar w)$ contains four variables. The problem of obtaining $K$ is thus reduced to that of obtaining the two-variable function $\tilde S_{a'}^{(a,b)}(x,y)$.

We now show that $\tilde S_{a'}^{(a,b)}(x,y)$ satisfies a system of coupled second-order partial differential equations (PDEs). The $k_\beta$ in~\eqref{eq:Si} obeys $\mathcal D_{\rho_z} k_{-2m-2a'}(z) = (m+a')(m+a'+1)k_{-2m-2a'}(z)$, with similar equations for the coordinates $\bar z$ and $\bar w$, \footnotemark[3].
By taking the differences, we obtain PDEs for $S_{a'}^{(a,b)}$, namely
$\left( \mathcal{D}_{\rho_z}-\mathcal{D}_{\bar\rho_z} \right)S_{a'}^{(a,b)}=0$ and 
$\left( \mathcal{D}_{\rho_z}-\mathcal{D}_{\bar\rho_w} \right)S_{a'}^{(a,b)}=0$.
After inserting~\eqref{eqstilde4} in these two PDEs, the system reduces to
\begin{subequations}\label{eq:PDE}
\begin{align}
 &\left[x^2(1-x)\partial_x^2-x^2\partial_x+y^2\partial_y^2+y\partial_y-x^2(1-y^2)\partial_x\partial_y/2\right.\notag\\
 &\quad \left.+\left(1/4-a^2-b^2\right)\right]\tilde S^{(a,b)}_{a'}=0\, ,\label{eq:PDE1}\\
&\left[\left(x^2(1-y)^2+4xy\right)\partial_x\partial_y-2y\partial_y-4ab\right]\tilde S^{(a,b)}_{a'}=0\,.\label{eq:PDE2}
\end{align}
\end{subequations}
To obtain an explicit expression for the bulk kernel, one must solve the PDE system. In~\cite{Carmi:2019cub}, solutions were obtained only in specific limits—namely, as $x \to 1$ and $y \to 0$ or $y \to \infty$—and not for general values of $a$, $b$, and $(x, y)$. In this work, we overcome this limitation and provide a solution valid in the full kinematic range.

\noindent \textit{Solving the PDE system}. 
Now, we derive a novel expression for the kernel with unequal scaling dimensions by solving~\eqref{eq:PDE}, which, as we will see, can be written in terms of the Appell functions---see Appendix~\ref{app:Appell}. Our strategy begins with solving the equation in specific, simple cases, and then inferring the general solution based on the insights gained from these special instances---a detailed step-by-step derivation is provided in Appendix~\ref{app:SolPDE}. We start by adding and subtracting Eqs. \eqref{eq:PDE1} and \eqref{eq:PDE2}. Afterward, we can put one of the two PDEs in its canonical form. Then one notices that there are two special cases, $b=1/2\mp a$, for which the PDEs are explicitly solvable. The solutions in these two cases are given by \footnote{Here $\tilde S^{(a,b)}= \tilde{S}^{(a,b)}_a+\tilde{S}^{(a,b)}_{-a}+\tilde{S}^{(b,a)}_b+\tilde{S}^{(b,a)}_{-b}$.}:
\begin{align}\label{eq:f2aaXX}
\tilde S^{(a,b=\frac{1}{2}-a)} &=-\frac{1}{2\pi}\ {}_2F_1\left(2a,1-2a,1;\xi\right) \,,\\ \nonumber
\tilde S^{(a,b=\frac{1}{2}+a)} &=-\frac{1}{2\pi}\ {}_2F_1\left(-2a,1+2a,1;\eta\right)\,,
\end{align}
where the $(\xi,\eta)$ coordinates are defined as $\xi\equiv\left(1 + y - \sqrt{4y/x + (1-y)^2}\right)/2$ and $\eta\equiv\left(1 + y - \sqrt{4y/x + (1-y)^2}\right)/(2y)$ \footnote{In terms of the variables $(w,\bar w, z, \bar z)$ or $(U',V',U,V)$, we have:
\begin{align}
\xi &\equiv \frac{1}{2}\left( 1 + y -\frac{s}{2}\right)=\frac{1}{2}\left( 1+\sqrt{V V'}+\frac{U(1-V')+U'(1-V)}{2\sqrt{U U'}}\right)\,, \quad \frac{\eta}{\xi+\eta-\xi \eta} =\frac{2}{1+y+\frac{s}{2}}=\frac{2}{1+\frac{q_1}{2}}\\
 \eta &\equiv \frac{1}{2y}\left( 1 + y -\frac{s}{2}\right)=\frac{1}{2\sqrt{V V'}}\left( 1+\sqrt{VV'}+\frac{U(1-V')+U'(1-V)}{2\sqrt{U U'}}\right)\,,\quad \frac{\xi}{\xi+\eta-\xi \eta} =\frac{2y}{1+y+\frac{s}{2}}= \frac{2}{1+p_2^{-1}} 
\end{align}
where $U'\equiv w\bar w$, $V'\equiv (1-w)(1-\bar w)$, $U \equiv z\bar z$, $V \equiv (1-z)(1-\bar z)$, and $s \equiv \sqrt{w\wb z\zb}\left(w^{-1}+\wb^{-1}+z^{-1}+\zb^{-1}-2\right)$, and $y=\sqrt{(1-z)(1-\bar z)(1-w)(1-\bar w)}$.}.

The special solutions \eqref{eq:f2aaXX} give us a hint that $(\xi,\eta)$ might be a natural set of coordinates. 
We thus transform the PDEs in~\eqref{eq:PDE} to the new variables $(\xi,\eta)$, and get
\begin{equation}\label{eq:dfjkdf}
\mathcal D_{(\eta,\xi)}^{(a,b)}\tilde S_{a'}(\xi ,\eta )
=0\,,\quad \mathcal D_{(\xi,\eta)}^{(a,-b)} \tilde S_{a'}(\xi,\eta)=0\,.
\end{equation}
where the introduced differential operator reads $\mathcal D_{(\eta,\xi)}^{(a,b)}\equiv\eta(1-\eta) \partial_\eta^2 +\xi \partial_\xi \partial_\eta +(1-2 \eta) \partial_\eta + \left((a+b)^2-1/4 \right)$. Remarkably, Eqs.~\eqref{eq:dfjkdf} are now explicitly in the form of the Appell function PDEs---see Appendix~\ref{app:Appell}. The solution to the PDEs is given in \eqref{eq:Stildeab}, in terms of the Appel $F_3$ function, this is
\begin{widetext}
\begin{equation}
\boxed{  \tilde S^{(a,b)}=-\frac{1}{2\pi}F_3(\xi,\eta) \equiv -\frac{1}{2\pi}F_3\left(\frac{1}{2}+a-b,\frac{1}{2}-a-b,\frac{1}{2}-a+b,\frac{1}{2}+a+b,1;\xi,\eta\right) \,.} \label{eq:Stildeab}
\end{equation}
\end{widetext}

Using~\eqref{eq:KKi}, the bulk kernel is obtained by action of the differential operator $\mathcal{D}_2$ (written in $x$ variables): 
\begin{equation}\label{eq:KBab}
K_B (z,\bar z,w,\bar w) = \frac{\mu_{(2)}}{2\pi} \frac{z\zb(w-\wb)sx^2}{8y^{\frac{3}{2}+a+b}}\partial_x F_3(\xi,\eta)\, ,
\end{equation}
 where $s \equiv \sqrt{w\wb z\zb}\left(w^{-1}+\wb^{-1}+z^{-1}+\zb^{-1}-2\right)$.  Taking the $x$ derivative, the final expression of $K_B$ is a sum of two $F_3$ functions with shifted parameters.

Expression \eqref{eq:Stildeab}---and, by extension \eqref{eq:KBab}---is our main result. The function $\tilde S$ and the bulk kernel $K_B$ of the DR, turn out to be an Appell $F_3$ function. The Appell functions are the simplest two-variable generalizations of the ${}_2F_1$ hypergeometric function. The remarkable simplicity of this result is notable. This result holds for and $d\geq 2$. It would be interesting to derive the $d=1$ dispersion kernel for mixed correlators~\cite{Paulos:2020zxx,Bonomi:2024lky,Carmi:2024tmp}.

\noindent \textit{Alternative expression for the kernel}. For practical numerical applications one would like to evaluate the $F_3(\xi,\eta)$ via its Taylor expansion around $\xi=\eta=0$. This series converges in the range $|\xi|,|\eta|<1$. However, in the DR~\eqref{eq:Gt} the variables run in the range $-\infty<\xi,\eta<0$, outside of the series convergence range. In order to overcome this, we will now obtain an alternative expression for the bulk kernel, which has a convergent series representation in the full range of the DR. The key is an identity that relates the Appell $F_3(\xi,\eta)$ function with a linear combination of Appell $F_2\left(\eta/(\eta+\xi-\xi\eta),\xi/(\eta+\xi-\xi\eta)\right)$ functions---see Appendix~\ref{app:Appell} for more details. Thus, instead of~\eqref{eq:Stildeab}, it will be useful for practical applications to use the following result for $\tilde S_{a}^{(a,b)}$ \footnote{Note that from~\eqref{eq:SabF2} one can get all four terms that go into~\eqref{eq:KKi}, using $\tilde S^{\left(a,b\right)}_{-a}= \tilde S^{\left(-a,-b\right)}_{a}$}:
\begin{widetext}
    \begin{align} 
 \tilde S^{\left(a,b\right)}_{a}=&\frac{
  \left(\cos (2 \pi  a)+\cos (2 \pi  b)\right)
 \sin \left(2\pi a\right)
 \Gamma_{2 a+1}
 \Gamma_{2b-2a}
 \Gamma_{-2b-2a}
 \Gamma_{\frac{1}{2}+a-b}
 \Gamma_{\frac{1}{2}+a+b}}
 {4 \pi ^4
 \left(\frac{\xi+\eta-\xi\eta}{\eta }\right)^{1+2 a}
 \left(\frac{\eta}{\xi}\right)^{\frac{1}{2}+a+b}} \nonumber \\
 &\times 
 F_2\left(1+2a;\frac{1}{2}+a-b,\frac{1}{2}+a+b;1+2a-2b, 1+2a+2b ;\frac{\eta }{\xi+\eta-\xi\eta},\frac{\xi}{\xi+\eta-\xi\eta}\right)\,.\label{eq:SabF2}
\end{align}
\end{widetext}

The $F_2$ function above  has a Taylor series expansion---see Appendix~\ref{app:Appell}, that converges in the range $0<\left|\eta/(\eta+\xi-\xi\eta)\right|+\left|\xi/(\eta+\xi-\xi\eta)\right|<1$. The DR has support in the range $-\infty <\xi,\eta <0$, and the Taylor series of the $F_2$ converges in this range. Thus, for numerical application of the DR, one can use the expression~\eqref{eq:SabF2}.
\vspace{0.1 cm}

\noindent \textit{Special cases and numerical checks}. Having obtained the full kernel of the conformal DR, we would now like to perform some checks.
In the case $b=0$ and arbitrary $a$, the Appell function in~\eqref{eq:Stildeab} reduces to a hypergeometric function,
\begin{equation}\label{eq:Stildea0}
\tilde S^{(a,0)}=-\frac{x^{a+\frac{1}{2}}}{2\pi} \, _2F_1\left(a+\frac{1}{2},a+\frac{1}{2};1;1-x\right)\,,
\end{equation}
where we used identities displayed in Appendix~\ref{app:Appell}. In the case $a=b=0$, we get $\tilde S^{(0,0)}= -x^{1/2}/(2\pi)\,{}_2F_1\left(1/2,1/2;1;1-x\right)$
reproducing the bulk kernel obtained in \href{https://link.springer.com/content/pdf/10.1007/JHEP09(2020)009.pdf#equation.3.23}{3.23}) of~\cite{Carmi:2019cub}. Furthermore, for the family of cases with $b=1/2-a +n$ with $n$ being a non-negative integer, the result is given in terms of associated Legendre functions.

\noindent \underline{Numerical checks}: Recalling the conformal DR~\eqref{eq:Gt}, we test our expression for the bulk kernel numerically by applying it to test correlation functions with the structure of generalized free-field correlators, i.e.,
\begin{equation} \label{eq:gff}
 \mathcal{G}_\text{GFF}(z, \zb)
 = \left(z \zb\right)^{r_1} \left[\left(1-z\right)\left(1-\zb\right)\right]^{r_2} = U^{r_1}V^{r_2}\,,  
\end{equation}
where $r_{1,2}$ are parameters that we are free to choose. Taking the $t$-channel double discontinuity, we obtain
$\dDisct{\mathcal{G}_\text{GFF}}=2{U}^{r_{1}}{V}^{r_{2}}\sin\left(\pi r_{2}\right)\sin\left[\pi\left(a+b+r_{2}\right)\right]$, which we plug into~\eqref{eq:Gt}---see Appendix~\ref{app:gff}. Using the bulk kernel arising from the Appell $F_2$ function of~\eqref{eq:SabF2}, we numerically perform the $w$ and $\bar w$ integrals. The integrals converge provided that the values of $r_1$ and $r_2$ are inside a specific finite range.  For many values of the parameters $(a,b,r_1,r_2)$, we compare the right-hand side and left-hand side of~\eqref{eq:Gt}, and obtain perfect matching in each case. These are highly nontrivial checks, implying the correctness of the DR.

\noindent \textit{Mellin-space dispersion relation}. In this Section we will establish that our position-space DR~\eqref{eq:Gt} is equivalent to a Cauchy-type DR,~\eqref{eq:cauchy}, in Mellin-space. In other words, the two DRs are related by a (double) Mellin transform. A similar equivalency was shown in~\cite{Caron-Huot:2020adz} for the case of equal scaling dimensions $a=b=0$, and we extend this to the mixed-correlator case.

Consider the four-point correlator in position space $\mathcal{G}\left(U,V\right)=\mathcal{G}^t\left(U,V\right)+\mathcal{G}^u\left(U,V\right)$. Conformal correlators have a useful representation in Mellin-space~\cite{Mack:2009mi,Mack:2009gy,Penedones:2010ue}, in terms of the Mellin amplitude $M\left(\rt,\ru\right)= M^t\left(\rt,\ru\right) + M^u\left(\rt,\ru\right)$, 
\begin{equation}\label{eq:MellinTrans}
    \mathcal{G}^{t}\left(U,V\right)=\int\frac{\diff\rt\diff\ru}{\left(4\pi \iu\right)^2}\Gamma^{\rt,\ru}_{\Delta_i}U^{\frac{\Delta-\rt-\ru}{2}}V^{\frac{\rt-\Delta_2-\Delta_3}{2}}M^{t}\left(\rt,\ru\right)\,,
\end{equation}
where $\Delta=\Delta_1+\Delta_2+\Delta_3+\Delta_4$, the product of gamma functions $\Gamma^{\rt,\ru}_{\Delta_i}$ is defined in \footnote{
    $\Gamma^{\rt,\ru}_{\Delta_i}\equiv
    \Gamma_{\frac{\Delta _1+\Delta _3-\rt}{2}} 
    \Gamma_{\frac{\Delta _2+\Delta _4-\rt}{2}} 
    \Gamma_{\frac{\Delta _2+\Delta _3-\ru}{2}}
    \Gamma_{\frac{\Delta _1+\Delta _4-\ru}{2} }
    \Gamma_{\frac{\rt+\ru-\Delta _1-\Delta _2}{2}}
    \Gamma_{\frac{\rt+\ru-\Delta _3-\Delta_4}{2}}$.
},  and $\rt$ and $\ru$ are analogs of the Mandelstam variables. The Mellin amplitude shares many properties with flat-space scattering amplitudes, in particular it obeys a single-variable Cauchy-type DR~\cite{Penedones:2019tng,Carmi:2020ekr,Caron-Huot:2020adz,Trinh:2021mll}, similar to~\eqref{eq:cauchy}:
\begin{align}\label{eq:439}
    M^t\left(\rt,\ru\right)\ =\int_{\rt+\epsilon-\iu\infty}^{\rt+\epsilon+\iu\infty}\frac{\diff\rt'}{2\pi \iu}\frac{M^t\left(\rt',\ru'\right)}{\rt-\rt'}\,.
\end{align}

A similar relation holds for $M^u\left(\rt,\ru\right)$, and we have that $\rt+\ru=\rt'+\ru'$. Combining expressions \eqref{eq:MellinTrans} and \eqref{eq:439} gives:
\begin{equation}\label{eq:MellinTrans2}
    \mathcal{G}^{t} =\int\frac{\diff\rt \diff\ru \diff\rt'}{4\left(2\pi \iu\right)^3}\Gamma^{\rt,\ru}_{\Delta_i}U^{\frac{\Delta-\rt-\ru}{2}}V^{\frac{\rt-\Delta_2-\Delta_3}{2}}\frac{M^{t}\left(\rt',\ru'\right)}{\rt-\rt'}\,.
\end{equation}

Following a procedure similar to that done in~\cite{Caron-Huot:2020adz}, we write the $t$-channel Mellin amplitude as an integral over the double discontinuity $\text{dDisc}_{t}[\mathcal{G}(U,V)]$,
\begin{align}
    &M^t(\rt',\ru') = \int \frac{\diff U' \diff V' U^{\prime\frac{\rt'+\ru'-\Delta}{2}}V^{\prime\frac{\Delta_2+\Delta_3-\rt'}{2}}}
    {U' V'\Gamma^{\rt',\ru'}_{\Delta_i}}\nonumber \\
&\times \frac{\text{dDisc}_{t}[\mathcal{G}(U',V')]}{ 2\sin\left(\tfrac{\pi\left(\rt'-\Delta_1-\Delta_3\right)}{2}\right) \sin\left(\tfrac{\pi\left(\rt'-\Delta_2-\Delta_4\right)}{2}\right)}
.\label{eq:Msdisc}
\end{align}
Now we insert~\eqref{eq:Msdisc} into~\eqref{eq:MellinTrans2}, and swap the order of the position and Mellin-space integrations. This now has the form of a position-space DR
\begin{equation}\label{eq:MellinDR}
    \mathcal{G}^{t}\left(U,V\right)=\int \diff U' \diff V'K_{\text{Mellin}}\text{dDisc}_{t}[\mathcal{G}(U',V')]\,,
\end{equation}
with a kernel $K_{\text{Mellin}}\equiv K_{\text{Mellin}}(U,V,U',V')$ given by:
\begin{widetext}
\begin{equation}\label{eq:Kmellin}
K_{\text{Mellin}}(U,V,U',V')
= \int \frac{\diff \rt \, \diff \ru \, \diff \rt'}{U'V'(2\pi \iu)^3} \frac{U^{\frac{\Delta-\rt-\ru}{2}}V^{\frac{\rt-\Delta_2-\Delta_3}{2}}}
{U^{\prime\frac{\Delta-\rt'-\ru'}{2}}V^{\prime\frac{\rt'-\Delta_2-\Delta_3}{2}}}
\frac{1}{\rt-\rt'}
\frac{\Gamma^{\rt,\ru}_{\Delta_i}/\Gamma^{\rt',\ru'}_{\Delta_i}}
{8\sin\left(\tfrac{\pi\left(\rt'-\Delta_1-\Delta_3\right)}{2}\right) \sin\left(\tfrac{\pi\left(\rt'-\Delta_2-\Delta_4\right)}{2}\right)}\,.
\end{equation}
\end{widetext}

Now, we compute~\eqref{eq:Kmellin} using the residue theorem. First we close the integration contour over $\rt'$ to the left and then we close the integration contour of $\rt$ and $\ru$ to the right. This leaves us with four towers of poles,
\begin{align}\label{eq:4393}
    K_{\text{Mellin}}= H_{a}^{(a,b)}+H_{-a}^{(a,b)}+H_{b}^{(b,a)}+H_{-b}^{(b,a)}\,,
\end{align}
where each tower of poles is given by \footnote{
\begin{equation}
\begin{split}
H_{a'}^{(a,b)}=&\sum _{\rt,\ru=0}^{\infty}
\frac{(-1)^{\rt+\ru+1} \Gamma_{-\frac{b a'}{a}+a'+1} \Gamma_{\frac{ba'}{a}-a'-\rt} \Gamma_{-\frac{b a'}{a}+u+a'}}
{2 \pi   \Gamma_{\rt+1} \Gamma_{\ru+1} \Gamma_{-u-2 a'-1} \Gamma_{\frac{b a'-a \left(\ru+a'+1\right)}{a}}
\sin \left(\frac{\pi  (a-b) a'}{a}\right)}\\
&\times \frac{{}_4\tilde{F}_3 \left(1,-\frac{b a'}{a}+a'+1,u+2 a'+2,u+a'-\frac{b a'}{a}+2;-\frac{b a'}{a}+a'+2,1-\rt, -\rt+a'-\frac{b a'}{a}+1;\frac{1}{v'} \right)}{u^{a'+\rt+\ru}  u^{\prime-a'-\rt-\ru+1}v^{\frac{(a+b) \left(a-a'\right)}{2 a}-u} v^{\prime\frac{1}{2} \left(-\frac{b a'}{a}+3 a'-a-b+2 \ru+4\right)}}
\end{split}
\end{equation}
where ${}_4\tilde F_3$ is the regularized ${}_4F_3$ function.}. We numerically compute the right-hand side of~\eqref{eq:4393}, and find perfect matching with our position-space bulk kernel given in~\eqref{eq:KBab}. In Appendix~\ref{app:coll} we contextualize this result with the one derived in the collinear limit in \cite{Trinh:2021mll}

\noindent \textit{Discussion}. In this work, we derived a conformal DR for mixed correlators. This new entry in the successful conformal bootstrap program enables the application of dispersive techniques to a broader class of CFT observables. It thus offers a flexible analytic tool with potential impact across CFT and AdS/CFT contexts.

Among the possible immediate applications, one could construct dispersive sum rules using either the position-space or Mellin-space representation of the conformal DR, and apply them to bootstrap the conformal data of mixed correlators in specific theories, such as $\mathcal{N}=4$ super Yang–Mills or the 3$d$ Ising model. In the $\mathcal{N}=4$ case, a numerical analysis using dispersive functionals was done in \cite{Caron-Huot:2022sdy,Caron-Huot:2024tzr}, obtaining two-sided bounds on the OPE coefficients of the Konishi operator as well as on the correlator itself. It would be intriguing to push this program forward and study bounds on other observables and models.

It should also be possible to derive the mixed-correlator conformal DR via a contour deformation argument, and to formulate DRs with subtractions \footnote{For the case of identical operators, such results were obtained in~\cite{Carmi:2019cub} and~\cite{Caron-Huot:2020adz}.}. Such a contour deformation proof could facilitate to generalize the DR to higher-point correlation functions. The four-point (and putative higher-point) conformal DRs can be viewed as novel mathematical theorems in complex analysis of multivariable functions. Taking a flat-space limit of the correlator DRs could result in new derivations of S-matrix DRs, which themselves play a significant role in the S-matrix bootstrap program. It would be interesting to derive the full conformal DR in the special case of one-dimensional CFTs, as well as for two-point functions in defect or boundary CFTs. Finally, a connection to thermal correlator DRs could follow by taking the heavy-heavy-light-light limit of our conformal DR. We leave these applications for future work.

\begin{acknowledgments}
\textit{Acknowledgments:} The work of D.C. and S.S. is supported by the Israeli Science Foundation (ISF) Grant No. 1487/21 and by the MOST NSF/BSF physics grant number 2022726. The work of S.S. is also partially supported by a PhD fellowship from the Israel Scholarship Education Foundation. The work of J.M. is partially supported by the ISF, Grant No. 1487/21 and by ANID FONDECYT Postdoctorado Grant No. 3230626.
\end{acknowledgments}

\onecolumngrid 
\appendix

\section{Appell functions}\label{app:Appell}
Here we review the Appell functions  $F_2$  and  $F_3$, along with key identities that relate them. See e.g. \cite{SpecialFunctions} for more details.
\subsection{Definition of \texorpdfstring{$F_2$ and $F_3$}{F2 and F3}}
The power series of the Appell $F_2$ function around $x=y=0$ is:
\begin{equation}\label{eq:F2def}
F_2(\alpha,\beta_1,\beta_2;\gamma_1,\gamma_2;x,y)=\sum_{i,j=0}^\infty\frac{(\alpha)_{i+j}(\beta_1)_{i}(\beta_2)_{j}}{(\gamma_1)_{i}(\gamma_2)_{j}i!j!}x^iy^j\, ,
\end{equation}
where $(\alpha)_i=\frac{\Gamma_{\alpha+i}}{\Gamma_{\alpha}}$ is the Pochhammer symbol, and the series converges for $|x|+|y|<1$.  The Appell $F_3$ power series, which converges  for $|x|<1$, $|y|<1$, is
\begin{equation}\label{eq:F3def}
F_3(\alpha_1,\alpha_2,\beta_1,\beta_2;\gamma;x,y)=\sum_{i,j=0}^\infty\frac{(\alpha_1)_{i}(\alpha_2)_{j}(\beta_1)_{i}(\beta_2)_{j}}{(\gamma)_{i+j}i!j!}x^iy^j\, .
\end{equation}
From this expression, it can be checked that $F_3$ satisfies the following coupled partial differential equations:
\begin{subequations}\label{eq:F3PDE}
\begin{align}
x(1-x)\frac{{\partial}^{2}{F_{3}}}{{\partial x}^{2}}+y\frac{\,{\partial}^{2}{F%
_{3}}}{\,\partial x\,\partial y}+\left(\gamma-(\alpha_1+\beta_1+1)x\right)\frac{%
\partial{F_{3}}}{\partial x}-\alpha_1\beta_1{F_{3}}&=0\,,\\
y(1-y)\frac{{\partial}^{2}{F_{3}}}{{\partial y}^{2}}+x\frac{\,{\partial}^{2}{F%
_{3}}}{\,\partial x\,\partial y}+\left(\gamma-(\alpha_2+\beta_2+%
1)y\right)\frac{\partial{F_{3}}}{\partial y}-\alpha_2\beta_2{F_{%
3}}&=0\,.
\end{align}
\end{subequations}

\subsection{Identities involving Appell \texorpdfstring{$F_2$, $F_3$}{F2, F3}}
A useful reduction formula for the $F_3$ function into ${}_{2}F_1$ is the following
\cite{SpecialFunctions}
\begin{equation}
\label{eq:dflsds}
    {F_{3}}\left(\alpha,\gamma-\alpha,\beta,\gamma-\beta,\gamma;x,y\right)=(1-y)^{%
\alpha+\beta-\gamma}{{}_{2}F_{1}}\left({\alpha,\beta,\gamma},x+y-xy\right)\,.
\end{equation}
This identity was used in \eqref{eq:Stildea0} to get the bulk kernel in the case $b=0$. The Appell $F_2$ and $F_3$ functions are related through the following identity \cite{SpecialFunctions}
\begin{equation}\label{eq:F2toF3}\begin{split}
    F_3\left(\alpha ,\alpha ',\beta ,\beta ',\gamma ,x,y\right)&
    =(-x)^{-\alpha } (-y)^{-\alpha '} f_{\left(\alpha ,\alpha ',\beta ,\beta ' \right)} F_2\left(\alpha -\gamma +\alpha '+1;\alpha ,\alpha ';\alpha -\beta +1,\alpha '-\beta '+1;\frac{1}{x},\frac{1}{y}\right)\\ &
    +(-x)^{-\alpha } (-y)^{-\beta '} f_{\left(\alpha ,\beta ',\beta ,\alpha ' \right)} F_2\left(\alpha -\gamma +\beta '+1;\alpha ,\beta ';\alpha -\beta +1,-\alpha '+\beta '+1;\frac{1}{x},\frac{1}{y}\right)\\ &
    +(-x)^{-\beta } (-y)^{-\alpha '} f_{\left(\beta ,\alpha ',\alpha ,\beta ', \right)} F_2\left(\beta -\gamma +\alpha '+1;\beta ,\alpha ';-\alpha +\beta +1,\alpha '-\beta '+1;\frac{1}{x},\frac{1}{y}\right)\\ &
    +(-x)^{-\beta } (-y)^{-\beta '} f_{\left(\beta ,\beta ',\alpha ,\alpha ' \right)} F_2\left(\beta -\gamma +\beta '+1;\beta ,\beta ';-\alpha +\beta +1,-\alpha '+\beta '+1;\frac{1}{x},\frac{1}{y}\right)\,,
\end{split}\end{equation}
where $\arg \left(-x\right)<\pi$ and $\arg \left(-y\right)<\pi$, and we defined $f_{(\lambda ,\mu ,\rho ,\sigma  )}=\frac{\Gamma (\gamma ) \Gamma (\rho -\lambda ) \Gamma (\sigma -\mu )}{\Gamma (\rho ) \Gamma (\sigma ) \Gamma (\gamma -\lambda -\mu )}$. Additionally, the Appell $F_2$  function obeys the transformation \cite{ANANTHANARAYAN2023108589}:
\begin{equation}\label{eq:F2imp}
F_2\left(\alpha ;\beta ,\beta ';\gamma ,\gamma ';x,y\right)=(1-x-y)^{-\alpha } F_2\left(\alpha ;\gamma -\beta ,\gamma '-\beta ';\gamma ,\gamma ';\frac{-x}{1-x-y},\frac{-y}{1-x-y}\right)\,.
\end{equation}
In order to transform the bulk kernel expression in \eqref{eq:Stildeab} into the expression written in \eqref{eq:SabF2}, we used the transformation of \eqref{eq:F2toF3} followed by \eqref{eq:F2imp}. By taking the derivative of \eqref{eq:SabF2}, we get the following result for $K^{\left(a,b\right)}_{a'}$:
\begin{equation}\begin{split}     K^{\left(a,b\right)}_{a'}=&\frac{\left(\bar{w}-w\right)z\bar{z}\mu_{(2)}}{8 \pi  \xi  \Gamma \left(-2 a'\right)} \Gamma_{2 b-2 a'}\Gamma_{-2 b-2 a'} \left(\frac{\xi+\eta-\xi\eta}{\eta }\right)^{-1-2 a'} \left(\frac{\xi }{\eta }\right)^{\left(1+\frac{b}{a}\right) a'-a-b}
     \\ \times &
     \left(\frac{\left(\frac{1}{4}-(a-b)^2\right)
     F_2\left(1+2 a';\left(1-\frac{b}{a}\right) a'-\frac{1}{2},\left(1+\frac{b}{a}\right) a'+\frac{1}{2};1+2 \left(1-\frac{b}{a}\right) a',1+2 \left(1+\frac{b}{a}\right) a';\frac{\eta }{\xi+\eta-\xi\eta},\frac{\xi }{\xi+\eta-\xi\eta}\right)}
     {\Gamma_{\frac{1}{2}-\left(1+\frac{b}{a}\right) a'}\Gamma_{\frac{3}{2}-\left(1-\frac{b}{a}\right) a'}}\right.
     \\ + &
     \left.\frac{\left(\frac{1}{4}-(a+b)^2\right)
     F_2\left(1+2 a';\left(1-\frac{b}{a}\right) a'+\frac{1}{2},\left(1+\frac{b}{a}\right) a'-\frac{1}{2};1+2 \left(1-\frac{b}{a}\right) a',1+2 \left(1+\frac{b}{a}\right) a';\frac{\eta }{\xi+\eta-\xi\eta},\frac{\xi }{\xi+\eta-\xi\eta}\right)}
     {\Gamma_{\frac{1}{2}-\left(1-\frac{b}{a}\right) a'} \Gamma_{\frac{3}{2}-\left(1+\frac{b}{a}\right) a'}}\right)
     .
\end{split}\end{equation}
See \eqref{eq:KKi}.

\section{Solving the partial differential equations system \texorpdfstring{\eqref{eq:PDE}}{(\ref*{eq:PDE})}}\label{app:SolPDE} 
A general linear second-order PDE for some function $f(x,y)$ can be written as:
\begin{equation}
a\left(x,y\right)\partial^2_{x}f+2b\left(x,y\right)\partial_x \partial_y f+c\left(x,y\right)\partial^2_y f+d\left(x,y\right)\partial_x f+e\left(x,y\right)\partial_y f+h\left(x,y\right)f=g\left(x,y\right)\,.
\end{equation}
This PDE is hyperbolic at a point $\left(x,y\right)$ if $b^2-ac>0$, and there exist a coordinate transformation $(x,y)\to (q,p)$ that can bring it to canonical form:
\begin{equation}
\partial_q \partial_p F+\alpha \left(q,p\right)\partial_q F+\beta \left(q,p\right)\partial_p F +\gamma\left(q,p\right)F=\epsilon \left(q,p\right)\,.
\end{equation}
The canonical variables $q(x,y)$ and $p(x,y)$ satisfy the equations:
\begin{subequations}
\begin{align}
a \partial_x q+\left(b+\sqrt{b^2-a c}\right)\partial_y q&=0\,,\\
a \partial_x p+\left(b-\sqrt{b^2-a c}\right)\partial_y p&=0\,.
\end{align}
\end{subequations}
For example, the wave equation $\partial_x^2 f-\partial_y^2 f=0$ reduces to $\partial{q} \partial{p}f=0$. The canonical coordinates are $q=x+y$ and $p=x-y$, and the solution to the PDE is $f=f_1\left(q\right)+f_2\left(p\right)=f_1\left(x+y\right)+f_2\left(x-y\right)$.\\

We are interested in solving the partial differential equations (PDE) system given in \eqref{eq:PDE}. We show below how to manipulate them into the PDE system for the Appell $F_3$ function. To that end, we first add and subtract equations $\eqref{eq:PDE1}\pm \frac{1}{2}\eqref{eq:PDE2}$:
\begin{subequations}\label{eq:PDEpm}
\begin{align}
&\left[ (1-x) x^2 \partial_x^2 +y^2 \partial_y^2-x^2 \partial_x+x y (x y-x+2) \partial_x \partial_y +\left(1/4-(a+b)^2\right) \right] f(x,y) =0\, ,\label{eq:PDEpm1}\\
&\left[ (1-x) x^2 \partial_x^2 +y^2 \partial_y^2-x^2 \partial_x+2y\partial_y+x (x y-x-2y) \partial_x \partial_y +\left(1/4-(a-b)^2\right) \right] f(x,y) =0\,,\label{eq:PDEpm2}
\end{align}
\end{subequations}
For each equation in \eqref{eq:PDEpm} we have a canonical form with different canonical coordinates. Choosing to put \eqref{eq:PDEpm1} in canonical form, gives the canonical coordinates $q_1 =2\sqrt{4\frac{y}{x}+\left(1-y\right)^2}+2y$, $p_1=2\sqrt{4\frac{y}{x}+\left(1-y\right)^2}-2y$. On the other hand, choosing to put \eqref{eq:PDEpm2} in canonical form, we get $q_2 =\frac{y}{\sqrt{4\frac{y}{x}+\left(1-y\right)^2}-1}$, $p_2=\frac{y}{\sqrt{4\frac{y}{x}+\left(1-y\right)^2}+1}$. We will use these coordinates below.

\subsection{Solving the case \texorpdfstring{$b=\frac{1}{2}-a$}{b=1/2-a}}

Equations \eqref{eq:PDEpm} in $(q_1,p_1)$ coordinates become:
\begin{subequations}\label{eq:PDExe1}
\begin{align}
0=&\left[\left(1/4-(a+b)^2\right)- \left(p _1-q _1\right){}^2 \partial_{p_1}\partial_{q_1} \right] f\left(q_1,p_1\right) \, ,\label{eq:PDExe1a}\\
0=&\left[(1/4-(a-b)^2) +2p _1 \partial_{p_1} +\left(p_1^2-4\right) \partial_{p_1}^2+2 q _1 \partial_{q_1 }
+2 \left(p_1 q_1-4\right) \partial_{p_1}\partial_{q_1} +\left(q_1^2-4\right) \partial_{q_1}^2   \right] f\left(q_1,p_1\right)\, .\label{eq:PDExe1b}
\end{align}
\end{subequations}
Expression \eqref{eq:PDExe1a} is now in canonical form, and we see that when $b=\frac{1}{2}-a$, then \eqref{eq:PDExe1a}, simplifies to a wave equation $f^{(1,1)}\left(q_1,p_1\right)=0$. The general solution to this wave equation is $f\left(q_1,p_1\right)=g\left(q_1\right)+h\left(p_1\right)$, which when substituted into \eqref{eq:PDExe1b} gives:   
\begin{equation}
    2 a (1-2 a) (g(q_1 )+h(p_1 ))+\left(q_1 ^2-4\right) g''(q_1 )+2 q_1  g'(q_1 )+\left(p_1 ^2-4\right) h''(p_1 )+2 p_1  h'(p_1 )=0\,.
\end{equation}
This separation of variables leads to independent ordinary differential equations (ODEs) for $g$ and $h$:
\begin{subequations}
\begin{align}
2 a (1-2 a) g(q_1 )+\left(q_1 ^2-4\right) g''(q_1 )+2 q_1  g'(q_1 )&=0\,,\\
2 a (1-2 a) h(p_1 )+\left(p_1 ^2-4\right) h''(p_1 )+2 p_1  h'(p_1 )&=0\,.
\end{align}
\end{subequations}
Since these are the ODEs of the Legendre functions, we have:
\begin{equation}
    f(q_1 ,p_1 )=c_1 P_{2 a-1}\left(\frac{q_1 }{2}\right)+c_2 Q_{2 a-1}\left(\frac{q_1 }{2}\right)+c_3 P_{2 a-1}\left(\frac{p_1 }{2}\right)+c_4 Q_{2 a-1}\left(\frac{p_1 }{2}\right)\,,
\end{equation}
Where $P_{2a-1}$ and $Q_{2a-1}$ are the first and second Legendre functions. We can determine the constants $c_i$ by comparing with with the series expression of the kernel in \eqref{eq:Si}. This gives $c_1=c_2=c_4=0$ and $c_3=-\frac{1}{2\pi}$. To summarize, the solution to \eqref{eq:PDE} in the case $b=\frac{1}{2}-a$ is:
\begin{equation}\label{eq:f2aa}
f(q_1 ,p_1 ) = -\frac{1}{2\pi} P_{2a-1}\left(\frac{p_1}{2}\right) =-\frac{1}{2\pi}\ {}_2F_1\left(2a,1-2a,1;\frac{1}{2}-\frac{p_1}{4}\right)\,, \quad b=\frac{1}{2}-a\,.
\end{equation}
where in the second equality we wrote the Legendre function as a ${}_2 F_1$.

We can further derive the result in the family of cases $b=\frac{1}{2}-a+n$, where $n$ is a non-negative integer:
\begin{align}\label{eq:sdjk8d}
  f(q_1 ,p_1 )=-\frac{\Gamma_{2a-2n}}{2^n \pi\Gamma\left(2a\right)}\left(-y\right)^{n+1}\left(\frac{1}{y}\frac{\partial}{\partial y}\right)^{n}\left(\left(4-\left(s-2y\right)^2\right)^{\frac{n}{2}}P_{2a-1-n}^{n}\left(\frac{s-2y}{2}\right)\right)\,, \quad  b=\frac{1}{2}-a+n\,,
\end{align}
where $P_\mu^\nu$ is the associated Legendre function and the partial derivative in $y$ is such that $s=2\sqrt{\frac{4y}{x}+(1-y)^2}$ is held fixed. $s$ was written below \eqref{eq:KBab}, in terms of the $(z,\bar z,w, \bar w)$ coordinates.

\subsection{Solving the case \texorpdfstring{$b=\frac{1}{2}+a$}{b=1/2+a}}

Equations \eqref{eq:PDEpm} in canonical $(q_2,p_2)$ coordinates become:
\begin{subequations}\label{eq:PDExe2}
\begin{align}
0=&[(1/4-(a+b)^2)-2 p_2^3 \partial_{p_2}+\left(1-p_2^2\right) p_2^2\partial_{p_2}^2+2 p_2 q_2 \left(1-p_2 q_2\right)\partial_{p_2}\partial_{q_2} -2 q_2^3\partial_{q_2}+\left(1-q_2^2\right)q_2^2\partial_{q_2}^2]f\left(q_2,p_2\right),\label{eq:PDExe2a}\\
0=&[(1/4-(a-b)^2) -2 \left(p_2-q_2\right){}^2 \partial_{p_2}\partial_{q_2}]f\left(q_2,p_2\right)\, .\label{eq:PDExe2b}
\end{align}
\end{subequations}
Expression \eqref{eq:PDExe2b} is now in canonical form, and we see that when $b=\frac{1}{2}+a$, then, \eqref{eq:PDExe2b} simplifies to a wave equation $f^{(1,1)}\left(q_2,p_2\right)=0$. Plugging this solution in \eqref{eq:PDExe2a}, and performing similar steps to the previous subsection, we obtain the solution:
\begin{equation}\label{eq:f2bb}
f(q_2,p_2) = -\frac{1}{2\pi} P_{2 a}\left(q_2^{-1}\right) =-\frac{1}{2\pi}\ {}_2F_1\left(-2a,1+2a,1;\frac{1}{2}-\frac{q_2^{-1}}{2}\right)\,, \quad b=a +\frac{1}{2}\,.
\end{equation}

\subsection{Solving for general \texorpdfstring{$a$ and $b$}{a and b}}

In \eqref{eq:f2aa} and \eqref{eq:f2bb} we solved for the cases $b=\frac{1}{2}\pm a$. The variables that entered the ${}_2 F_1$'s there are:
\begin{equation}
  \xi \equiv \frac{1}{2}-\frac{p_1}{4}=\frac{1}{2}+ \frac{y - \sqrt{\frac{4y}{x} + (1-y)^2}}{2}\,, \quad
\eta  \equiv \frac{1}{2}-\frac{q_2^{-1}}{2}=\frac{1}{2} +\frac{1 - \sqrt{\frac{4y}{x} + (1-y)^2}}{2y}\,.
\end{equation}
This motivates writing our PDEs in these new coordinates $(\xi,\eta)$. Transforming \eqref{eq:PDEpm} to the variables $(\xi,\eta)$, the PDEs become
\begin{subequations}\label{eq:PDExe}
\begin{align}
&\left[ \eta(1-\eta) \partial_\eta^2 +\xi \partial_\xi \partial_\eta +(1-2 \eta) \partial_\eta + \left((a+b)^2-1/4 \right) \right]f(\xi ,\eta )=0\,,\\
&\left[ \xi(1-\xi) \partial_\xi^2 +\eta \partial_\xi \partial_\eta +(1-2 \xi) \partial_\xi + \left((a-b)^2-1/4 \right) \right]f(\xi ,\eta )=0\,.
\end{align}
\end{subequations}
Remarkably, this is the PDE system satisfied by the Appell $F_3$ function, i.e., \eqref{eq:F3PDE}. The solution is
\begin{equation}
\label{eq:skjhd7}
f(\xi ,\eta )=-\frac{1}{2\pi}F_3\left(\frac{1}{2}+a-b,\frac{1}{2}-a-b,\frac{1}{2}-a+b,\frac{1}{2}+a+b,1;\xi,\eta\right).
\end{equation}
In the main text we used this result in \eqref{eq:Stildeab}. It is the result for the function $\tilde S^{(a,b)}$.

\section{Dispersion relation for generalized free fields}\label{app:gff}

Here we present explicit expressions on the dispersion relation for mixed correlators in the case of generalized free fields. Employing the cross ratios $U=z\zb$, $V=(1-z)(1-\zb)$, the correlator \eqref{eq:gff} is written simply as $\mathcal G_{\text{GFF}}(U,V)=U^{r_1}V^{r_2}$. The double discontinuities are given by
\begin{subequations}\label{eq:dDiscGFF}
\begin{align}\dDisct{\mathcal{G}_\text{GFF}\left(U,V\right)}&=2{U}^{r_{1}}{V}^{r_{2}}\sin\left(\pi r_{2}\right)\sin\left[\pi\left(a+b+r_{2}\right)\right]\, ,\\
\dDiscu{\mathcal{G}_\text{GFF}\left(U,V\right)}&=\dDisct{\mathcal{G}_\text{GFF}\left(\frac{U}{V},\frac{1}{V}\right)}=2{U}^{r_{1}}{V}^{-a-r_{1}-r_{2}}\sin\left[\pi\left(a+r_{1}+r_{2}\right)\right]\sin\left[\pi\left(b+r_{1}+r_{2}\right)\right]\,,
\end{align}
\end{subequations}

The full dispersion relation, including the $t$- and $u$-channels in $\rho$ coordinates reads
\begin{align}
\mathcal{G}\left(\rz,\rzb\right)=&\int\limits_{0}^{1}\diff\rwb\int\limits_{0}^{\rz\rzb\rwb}\diff\rw\frac{16\left(1-\rw\right)\left(1-\rwb\right)}{\left(1+\rw\right)^{3}\left(1+\rwb\right)^{3}}K_{B}^{\left(a,b\right)}\left(\rz,\rzb,\rw,\rwb\right)\dDisct{\mathcal{G}\left(\rw,\rwb\right)}\notag\\
-&\int\limits_{0}^{1}\diff\rwb\int\limits_{0}^{\rz\rzb\rwb}\diff\rw\frac{16\left(1-\rw\right)\left(1-\rwb\right)}{\left(1+\rw\right)^{3}\left(1+\rwb\right)^{3}}\left(\frac{\left(1+\rz\right)\left(1+\rzb\right)}{\left(1-\rz\right)\left(1-\rzb\right)}\right)^{2a}K_{B}^{\left(a,-b\right)}\left(-\rz,-\rzb,\rw,\rwb\right)\dDiscu{\mathcal{G}\left(\rw,\rwb\right)} \notag\\
+&\int\limits_{0}^{1}\diff\rwb\,K_{C}^{\left(a,b\right)}\left(\rz,\rzb,\rw,\rwb\right)\dDisct{\mathcal{G}\left(\rw,\rwb\right)}\notag\\
-&\int\limits_{0}^{1}\diff\rwb
\left(\frac{\left(1+\rz\right)\left(1+\rzb\right)}{\left(1-\rz\right)\left(1-\rzb\right)}\right)^{2a}K_{C}^{\left(a,-b\right)}\left(-\rz,-\rzb,\rw,\rwb\right)\dDiscu{\mathcal{G}\left(\rw,\rwb\right)}\,.\label{eq:intgff}
\end{align}
To particularize to case of generalized free fields, we consider the correlator \eqref{eq:gff} and the double discontinuity \eqref{eq:dDiscGFF}. As mentioned in the main text, the bulk kernel involves complicated expressions of Appell $F_2$ functions and thus we have to compute numerically the integrals appearing in this expression. Choosing parameters $a$, $b$, $r_1$ and $r_2$ in which the integrals converge, we get a perfect match  between both sides of \eqref{eq:intgff}.

\section{Collinear limit and relation to the results of \cite{Trinh:2021mll}}\label{app:coll}

The work of \cite{Trinh:2021mll} starts from a Mellin-space dispersion relation, afterwards transforming to position space. The result is an expression for the position-space dispersion kernel for mixed correlators in the collinear limit. We take the collinear limit of our kernel \eqref{eq:KBab}, and compare the result to Eqs.~(\href{https://link.springer.com/content/pdf/10.1007/JHEP03(2022)032.pdf#equation.3.18}{3.18}-\href{https://link.springer.com/content/pdf/10.1007/JHEP03(2022)032.pdf#equation.3.21}{3.21}) of \cite{Trinh:2021mll}. The two results match, but only after correcting a mistake in (\href{https://link.springer.com/content/pdf/10.1007/JHEP03(2022)032.pdf#equation.3.21}{3.21}) of \cite{Trinh:2021mll}.
The correct equation should be $\mathfrak{B}^{\mathfrak{a},\mathfrak{b}}_{12}=\left(u'/u\right)^{\mathfrak{a}+\mathfrak{b}} \left(v/v'\right)^{\mathfrak{a}-\mathfrak{b}}\mathfrak{B}^{-\mathfrak{a},-\mathfrak{b}}_{34}$.

Also note that in order to relate our results to those in \cite{Trinh:2021mll}, one should transform $x_2\leftrightarrow x_3$. This induces the relations $u=1/U$, $v=U/V$ and  $\mathfrak{a}=-a^{2\leftrightarrow3},\mathfrak{b}=b^{2\leftrightarrow3}$. Here $(\mathfrak{a}, \mathfrak{b},u,v)$ are the parameters in \cite{Trinh:2021mll}, and $(a,b,U,V)$ are the parameters in our work. Additionally, the relation between the $\mathcal G$ functions is $\mathfrak{G}\left(u,v\right)=U^\mathfrak{b}\mathcal{G}^{2\leftrightarrow3}\left(U,V\right)$. One obtains the position-space bulk kernel in the collinear limit $u\to 0$, with the result
$\mathfrak{K}(u \rightarrow 0)=\mathfrak{B}^{\mathfrak{a},\mathfrak{b}}_{12}+\mathfrak{B}^{\mathfrak{a},\mathfrak{b}}_{34}$ where
$\mathfrak{B}^{\mathfrak{a},\mathfrak{b}}_{34}=
 \frac{\Gamma_{\mathfrak{a}+\mathfrak{b}} \left(v+v'-u'\right) \, _2\tilde{F}_1\left(\mathfrak{a},\mathfrak{b};\mathfrak{a}+\mathfrak{b}-\frac{1}{2},-\frac{1}{\chi }\right)}
 {4 \pi ^{3/2} u^{\mathfrak{a}+\mathfrak{b}} v^{\frac{3}{2}-\mathfrak{a}}  v'^{\frac{3}{2}-\mathfrak{b}}\chi ^{\mathfrak{a}+\mathfrak{b}-\frac{3}{2}}}
$ and $\chi=\frac{4 v v'}{\left(v-\left(\sqrt{u'}+\sqrt{v'}\right)^2\right) \left(v-\left(\sqrt{u'}-\sqrt{v'}\right)^2\right)}$.


\onecolumngrid
\bibliography{2References.bib}
\bibliographystyle{apsrev4-2} 

\end{document}